\documentclass[conference, letter]{IEEEtran}

\usepackage{svg}
\usepackage[hidelinks]{hyperref}
\usepackage{cite}
\usepackage{amsmath,amssymb,amsfonts}
\usepackage{glossaries}
\usepackage{algorithmic}
\usepackage{graphicx}
\usepackage{textcomp}
\usepackage{xcolor}
\usepackage{soul}
\usepackage{flushend}

\newcommand{\orcid}[1]{\href{https://orcid.org/#1}{\includesvg[height = 2ex]{image/ORCIDiD_iconvector.svg}}}

\newacronym{mis}{MIS}{minimally-invasive surgery}
\newacronym{vr}{VR}{virtual reality}
\newacronym{ui}{UI}{user interface}
\newacronym{fem}{FEM}{finite element modelling}
\newacronym{fe}{FE}{finite elements}
\newacronym{fps}{FPS}{frames per second}
\newacronym{pbr}{PBR}{physically-based rendering}
\newacronym[plural=ODEs, firstplural=ordinary differential equations]{ode}{ODE}{ordinary differential equation}
\newacronym[plural=DoFs, firstplural=degrees of freedom]{dof}{DoF}{degree of freedom}
\newacronym[plural=GUIs, firstplural=graphical user interfaces]{gui}{GUI}{graphical user interface}

\begin{document}
\IEEEsettopmargin{t}{60pt}
\def\IEEEtitletopspace{16pt} 
\title{Filasofia: A Framework for Streamlined Development of Real-Time Surgical Simulations}

\author{\IEEEauthorblockN{Vladimir Poliakov \orcid{0000-0003-4179-4014},
Dzmitry Tsetserukou \orcid{0000-0001-8055-5345}, ~\textit{Member,~IEEE},\\ and
Emmanuel Vander Poorten \orcid{0000-0003-3764-9551},~\textit{Member,~IEEE}}}

\maketitle

\begin{abstract}
Virtual reality simulation 
has become a popular approach 
for training and assessing medical students. 
It offers diverse scenarios, realistic visuals, 
and quantitative performance metrics for objective evaluation. 
However, creating these simulations 
can be time-consuming and complex, 
even for experienced users. 
The SOFA framework is an open-source solution 
that efficiently simulates finite element (FE) models in real-time. 
Yet, some users find it challenging to navigate the software 
due to the numerous components required for a basic simulation 
and their variability. 
Additionally, SOFA has limited visual rendering capabilities, 
leading developers to integrate other software for high-quality visuals. 
To address these issues, we developed Filasofia, 
a dedicated framework that simplifies development, 
provides modern visualization, 
and allows fine-tuning using SOFA objects. 
Our experiments demonstrate that 
Filasofia outperforms conventional SOFA simulations, 
even with real-time subdivision. 
Our design approach aims to streamline development 
while offering flexibility for fine-tuning. 
Future work will focus on further 
simplification of the development process for users.
\end{abstract}


\section{Introduction}


Presently, it is hard to imagine modern-day medicine,
\gls{mis} in particular,
without elaborate training. 
While being beneficial for the patient, 
\gls{mis} introduces an additional level of complexity 
for the clinician 
due to the inversion of motion and limited workspace. 
Thus, it becomes vital 
to ensure rigorous training of future endoscopists 
to guarantee safety 
and an adequate success rate. 
One of the approaches 
to training and assessing medical students 
that have gained traction in recent years 
is \gls{vr} simulation. 
When properly implemented,
\gls{vr} simulation enables 
a higher level of variability of rendered scenarios, 
visual realism and, most importantly, 
a broad body of quantitative performance metrics 
that can be used for objective assessment of a trainee \cite{Gallagher2012}.
On the other hand, the design of such simulations 
is often time-consuming and not trivial
even for experienced users.
One of the open-source solutions 
that a developer can opt for is 
the SOFA framework, which enables real-time simulation
of \gls{fe} models with high efficiency \cite{FaureFrancois2012SAMF}.
Nonetheless, the entry barrier to this software
can still be daunting for some users 
simply due to the number of components
needed to run a basic simulation
and their variability.
Another factor that makes SOFA less appealing
as a solution for surgical simulation design  
is its somewhat limited visual rendering capabilities.
As such, developers spend more time
integrating alternative software to enable
high-quality visual rendering in SOFA.

\begin{figure}[t]
\centering
\includegraphics[width=0.4\textwidth]{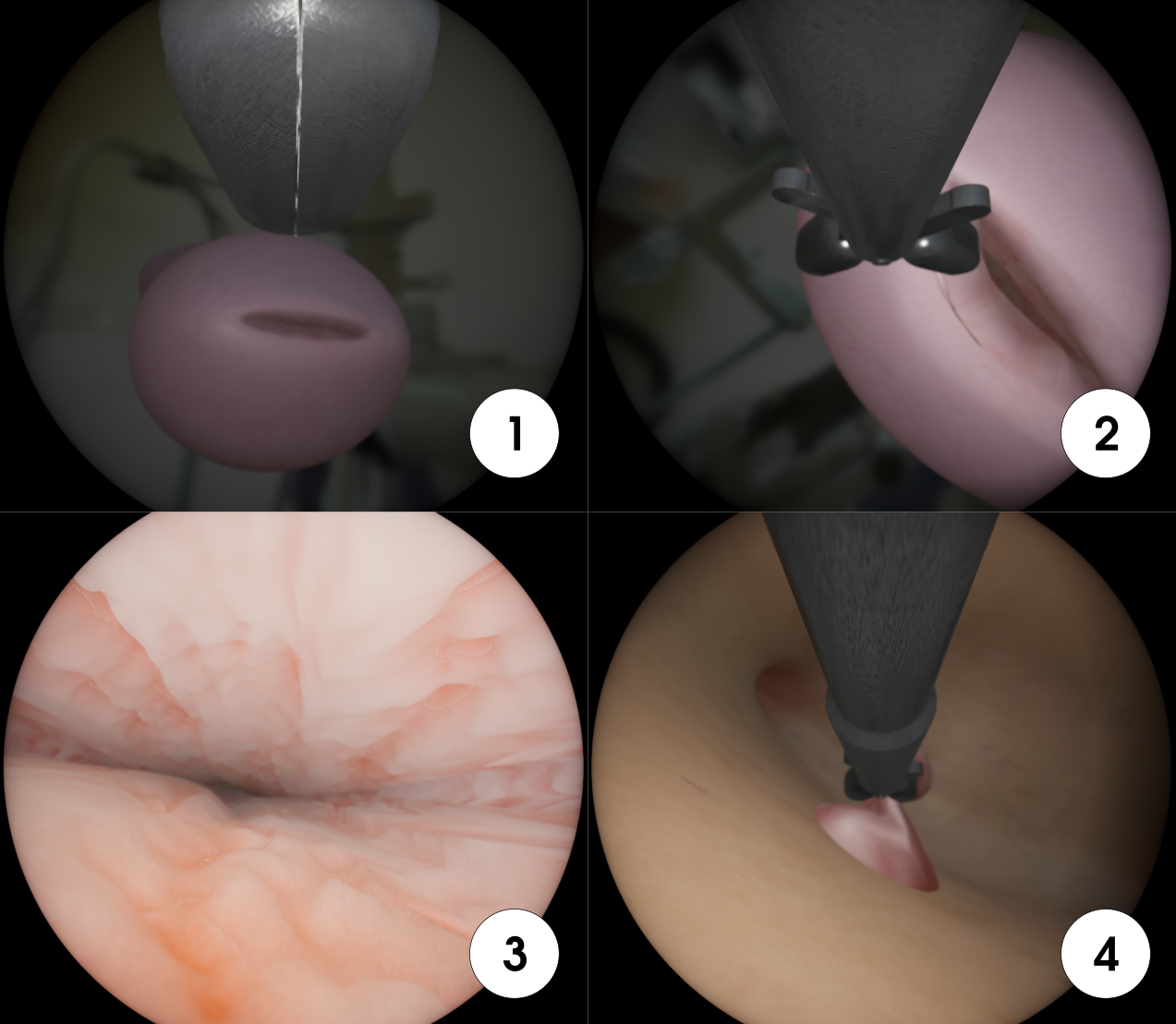}
\caption{A sample scene depiction 
of a hysteroscopic simulator developed using Filasofia:
(1)~the external os of the uterus, 
(2)~the dilation of the os with hysteroscopic forceps,
(3)~the cervical canal,
and (4)~the removal of a polyp.}
\label{fig:filasofia}
\end{figure}

To overcome these shortcomings,
we designed Filasofia\footnote{\href{https://gitlab.com/polyakovkrylo/filasofia}{https://gitlab.com/polyakovkrylo/filasofia}} (Figure~\ref{fig:filasofia}),
a dedicated framework
that simplifies the development process
for users and enables modern visualization,
while still providing full access 
to underlying SOFA objects
for fine-tuning.

\section{Overview of the framework}

\subsection{Software architecture}

\begin{figure*}[ht]
\centering
\includegraphics[width=0.8\textwidth]{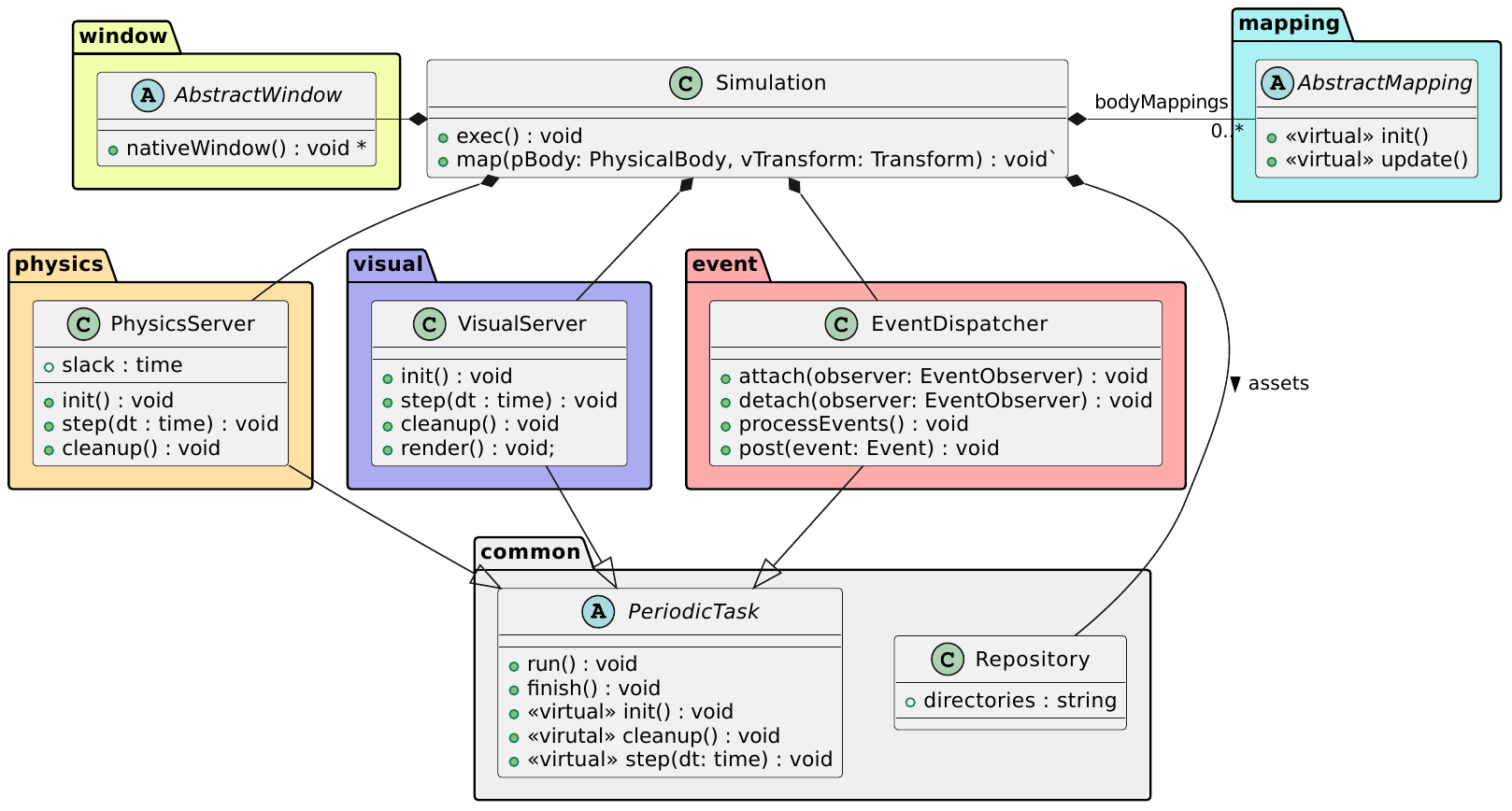}
\caption{The overview class diagram of the software architecture of Filasofia,
including the main components of the framework:
(1) the visual module, which is responsible for the graphical rendering,
(2) the physics module, which performs the physics simulation,
(3) the event module, which is responsible for the implementation 
of event-based scripting,
(4) the mapping module connecting physical bodies to their
representation in the visual domain,
and (5) the window module, which provides an abstraction layer
for window management and input access.
Note: some class properties and relations between classes
are omitted to improve the clarity of the diagram.}
\label{fig:software_architecture}
\end{figure*}

\textit{Filasofia} is a framework 
for surgical simulation development.
It has two main components at its core:
(1)~the physics server and 
(2)~the visual server (Figure \ref{fig:software_architecture}).
The physics server is responsible for 
collision detection and response computation,
integration and solving state equations, 
state updates and haptic rendering.
The physics server is primarily 
based on the SOFA framework.
Most of the physical body parameters
can be controlled at the time of the body creation.
The user can pass a configuration
that defines the type of the body,
solver parameters, 
e.g. precision and rayleigh coefficients;
collision model parameters,
e.g. type of primitives used,
detection proximity, and collision group.
For soft bodies, 
the user can define 
the type of the simulation model
(mass-spring of \gls{fem}),
and, depending on the model,
stiffness, Young's modulus,
and Poisson ratio.

The visual server renders 
visual entities,
such as visual bodies,
lights, cameras, and \glspl{ui}
on the screen. 
Additionally,
the visual server is also responsible
for the online subdivision of visual meshes.

There is an important distinction between a visual body and a physical body.
The visual body is a visual server entity 
that can be rendered on the screen,
while the physical body is an entity 
that is only employed in the physics simulation.
To represent a physical body in the visual domain,
one can create a corresponding visual body 
and map it to the physical body.
In this case, any state change in the physical body
will be propagated to the visual body 
with each update of the visual server.
There are several mapping schemes available in Filasofia.
The type of mapping is selected automatically
based on the type of the physical body
and its properties. 
It is also possible to map several visual entities
to one physical body.
This might be particularly useful when creating
articulated systems or creating objects
that encapsulate several visual entities,
e.g. a light source and a camera integrated into an endoscope.

The two servers are decoupled from each other
and run in parallel threads.
This allows them to have
different update rates
since the physics server typically 
has rigorous timing requirements
and requires a much higher update rate.
The only synchronisation point
between the servers is the mapping phase
when the states of the physical bodies
are mapped to their visual representations.

The window system contains 
yet another level of abstraction, 
which makes it possible to 
use different window manager frameworks.
Currently, Filasofia uses SDL~\cite{sdl}
to create windows and contexts.
If needed, it should be relatively easy
to also integrate other frameworks,
e.g. GLFW or Qt. 
All that is required from a derived class
is to provide a pointer to the created context.

Finally, to implement event-triggered behaviour,
Filasofia provides the event-observer mechanism,
which triggers a set of attached callbacks
every time a certain event occurs.
This is done in the \textit{EventDispatcher} class,
which acts as an event-managing task running in a dedicated thread.
Any object having access to the dispatcher
can post an event, custom or predefined.
Once posted, the event is added to the dispatcher's queue.
With each update,
the dispatcher will iterate over the events in the queue:
for each event in the queue, it will notify each observer 
that is attached to this type of event.
There are several predefined types of events 
(physics and visual server updates,
mouse events, key events, etc.)
along with four custom event types.
If needed, a developer can create 
their own type of event derived from the \textit{Event} class
and set its type to one of the \textit{CustomEventX} types.
Since an event is passed as an argument to its observers, 
it can also be extended to carry additional data
(timestamps, measurements, etc.).

\subsection{Physics server}

\begin{figure*}
\includegraphics[width=0.9\textwidth]{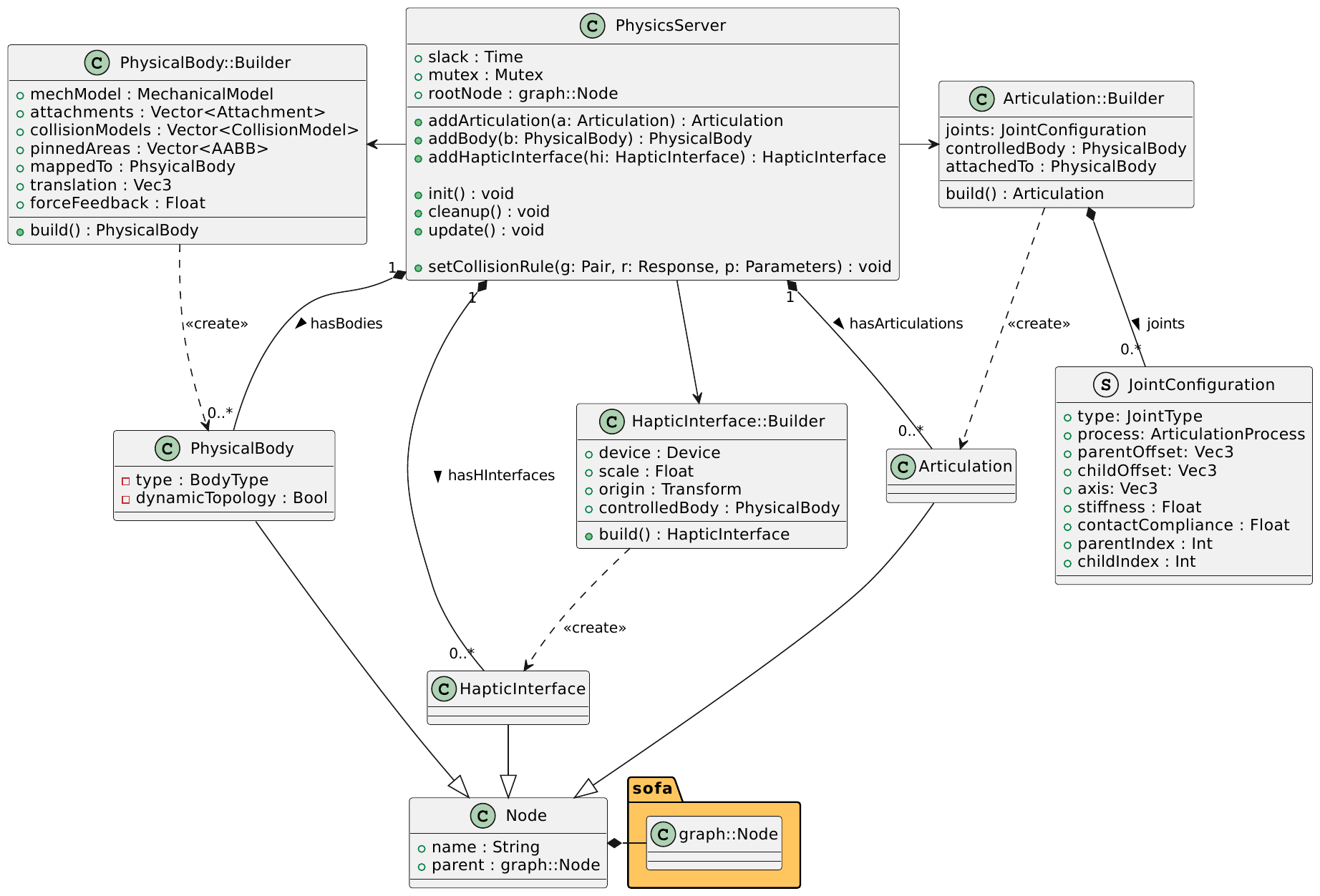}
\caption{The class diagram of the physics simulation component.
Note: some class properties and relations between classes
are omitted to improve the clarity of the diagram.}
\label{fig:physics_diagram}
\end{figure*}

The physics server (Figure~\ref{fig:physics_diagram}) is responsible for
the simulation of physical interaction 
between the bodies in the scene, including:
(1)~collision detection, 
(2)~collision response calculation,
(3)~integration of forces 
acting on each body in the simulation,
(4)~solving the resulting linear systems to compute 
the current state of those bodies
and~(5) haptic feedback calculation.
The physics server is based  on the SOFA framework \cite{FaureFrancois2012SAMF}.
SOFA is an efficient framework for physics-based simulation.
It allows developers to simulate interaction and deformation in real-time,
using multitudes of methods for each component of the simulation,
e.g. different force fields (mass-spring, \gls{fem}, hyperelastic \gls{fem}),
constraints (bilateral and unilateral), 
integration schemes (implicit and explicit) 
and linear system solvers (iterative and direct).
However, the number of components and their possible combinations
is so vast that it introduces a significant entry barrier for a novice user.
To address this problem, Filasofia automatically creates 
all necessary components to simulate a physical body 
based on its initial configuration.
To fine-tune each component,
a set of parameters can also be passed to the builder class.
A user can still opt to use direct access to the created node
representing the body in the SOFA domain and change it using the SOFA API,
which further increases the flexibility of the framework.

The physics server is a periodic task, which runs
in a dedicated thread and controls 
the creation and simulation of all physical entities.
To improve the stability of the physics simulation,
we introduced the slack parameter.
A real-time periodic task should be executed 
with a fixed period $T$ to guarantee the stability of the system.
In this case, the simulation timestep $dt$ is constant:

\begin{equation}
  dt=T=const
\end{equation}

However, in the cases of large computations due to 
e.g. high resolution of the model 
or weak performance of the computer,
the time required to compute one simulation step 
may become bigger than the period of the task,
leading to fluctiations in the timestep:

\begin{equation}
  dt=T+t_d
\end{equation}

where $t_d$ is the difference between the actual
and planned wake-up times.
While it is normal to 
have some fluctuations in the update rate of the physics loop,
exceeding a certain value of $dt$ may cause instability of simulation.
To overcome this issue,
the slack parameter can be used to limit the maximum value of $dt$:

\begin{equation}
  dt = min(T+t_d, T+t_s)
\end{equation}

where $t_s$ is the slack time allowed for the task.
This way, the simulation timestep will never exceed a predefined value,
leading to an occasionally less accurate, but stable solution.

There are three main types of physical entities in Filasofia:
(1)~a physical body,
(2)~a haptic interface,
and (3)~an articulation.

\subsubsection{Physical body}

A physical body is a nested entity that encapsulates all 
components required to simulate a body in SOFA.
Depending on the type, a physical body can be rigid or soft.
A rigid body is a set of 6-\gls{dof} elements,
each of which contains information about 
the position and orientation of a corresponding sub-element of that body.
Simple rigid bodies contain one 6-\gls{dof} element;
complex, e.g. articulated rigid bodies, 
can contain more than one 6-\gls{dof} element.

A soft body is a set of 3-\gls{dof} elements,
each of which defines the position of a corresponding node of that body.
The key feature of soft bodies is that they can deform.
There are two methods for deformation to choose from in Filasofia:
(1)~the mass-spring model, which is less accurate
but also less computationally demanding, 
and (2)~the tetrahedron \gls{fem} model,
which is more accurate but requires more time to compute.
In both cases, the user needs to provide a volumetric mesh
that defines the topology and geometrical configuration 
of the body.

Each physical body can contain a collision model.
The collision model of a body is defined 
by a triangular mesh 
and the following set of parameters:

\begin{enumerate}
  \item \textit{Primitives}: the type(s)
  of primitives to be used with the collision model 
  (points, lines, or triangles).
  \item \textit{Proximity}: the maximum distance 
  between a collision primitive 
  and the collision primitive of another body
  to detect intersection; if both bodies
  have a non-zero collision proximity,
  the total distance becomes equal 
  to the sum of their proximities.
  \item \textit{Group}: the exclusive collision
  group of the body 
  -- only collisions between bodies in the same group
  will be detected.
  \item \textit{Self-collision}: the flag controlling
  whether the collision primitives of this body
  can collide with each other.
  \item \textit{Cutting}: the flag controlling
  whether this collision model should be used
  for the registration of cuts. 
\end{enumerate}

Each physical body can have several collision
models, e.g. one model for cut registration
and another for collision simulation.
To do so, the user should repeatedly call 
the \textit{collisionModel()} function 
of the physical body builder
with different collision parameters
or different triangular meshes
-- each call will add a new collision model.

Each physical body can be attached to another physical body
or pinned to a point in space.
The attachment configuration parameters
passed to the physical body builder
contain stiffness of the attachment constraints,
indices of the body nodes that should be attached
and a flag controlling the activation 
of the barycentric attachment mode.
Barycentric attachment 
is used to connect bodies
when the number of attached nodes
is not equal to the number of nodes
they are attached to.
Barycentric attachment
does not connect nodes of two bodies directly
but rather creates a proxy model 
that clones the nodes of the attachable body
and maps it to the second body using 
the barycentric mapping algorithm.
These proxy nodes are then pinned
to the original body nodes
or attached to them using stiff springs, 
depending on the stiffness parameter 
in the attachment configuration.

To simulate state changes,
each body contains an \gls{ode} integrator
and linear solver components.
Currently, Filasofia supports two integration schemes:
explicit Euler and implicit Euler.
With the explicit Euler scheme,
the new state is computed based on
the information from the previous time step:

\begin{equation}
  x(t+dt) = x(t) + dt \cdot v(t)
\end{equation}

This approach is typically fast to solve
but prone to instabilities.
The explicit scheme is typically used to solve
non-stiff problems.
For stiff differential equations,
implicit solvers can be used.
The implicit solver computes the current state
based on the information of the current timestep:

\begin{equation}
  x(t+dt) = x(t) + \cdot v(t+dt)
\end{equation}

To solve the created linear system,
the user can use one of the solvers available in Filasofia:

\begin{enumerate}
  \item \textit{Direct solver}: 
  Direct solvers seek to find the exact solution to the system
  (\glspl{dof} at the next time step) in one iteration
  by computing the inverse matrix. 
  Various methods exist to compute the inverse matrix in SOFA.
  To avoid confusion for the user, Filasofia only uses 
  the Cholesky decomposition method~\cite{press2007numerical}.

  \item \textit{Iterative solver}:
  Iterative methods converge towards the solution gradually
  by approximating the solution more precisely with each step.
  The estimated error decreases with each iteration
  until it reaches the desired level 
  or the maximum number of iterations is reached.
  Filasofia uses the conjugate gradient method 
  as the iterative solver~\cite{shewchuk1994introduction}.
  
\end{enumerate}

By default, a body uses 
the implicit integration scheme 
with the conjugate gradient linear solver.

\subsubsection{Interactive topology modification}

At the creation time, each soft body 
can be configured to have a dynamic topology.
With the \textit{dynamicTopology} flag activated, a body can simulate
topology modification, 
including rupture and cutting simulation.
Since both rupture and cutting simulation
are not available in SOFA,
these features were developed as separate plugins for the SOFA framework
and then integrated into Filasofia.

The rupture simulation performs 
the separation of the body elements
when the stress level exceeds the \textit{tearingThreshold} value
defined in the mechanical parameters of the body builder.
The cutting simulation feature performs 
the registration of the cut path
and online remeshing of the body topology
according to the hybrid cutting simulation method~\cite{steinemann2006hybrid}. 

To propagate topological changes to the visual body,
a dedicated mapping algorithm is implemented.
This mapping scheme is chosen automatically
when the dynamic topology flag is set to true.
During the simulation, any topological changes
that occurred in the physical body will be propagated 
to the visual body too.

\subsubsection{Articulation}

The articulation class defines a link tree,
in which each element is a 1-\gls{dof} node that
can move linearly along a certain axis
or rotate around it. 
Each displacement of a parent node
will also cause displacement of successor nodes.
The articulation class is used to define
the articulation process of a rigid body.
The rigid body that should be controlled
with a given articulation object is passed 
as the \textit{controlledBody} property.

An articulation object can be configured
in the builder using the following parameters:

\begin{enumerate}
  \item \textit{Controlled body}: the body that 
  this articulation will control.
  \item \textit{Attachment body}: the body that 
  serves as the reference frame for this articulation
  \item \textit{Joint configuration}: 
  each joint in the articulation
  should be configured using 
  the \textit{JointConfiguration} structure, which contains:
   \begin{itemize}
     \item \textit{Joint type}: prismatic or revolute.
     \item \textit{Parent and child offsets}: offset 
     from the origin of the joint transformation to the parent/child node.
     \item \textit{Axis}: axis of rotation/translation.
     \item \textit{Parent and child indices}: 
     indices of the parent and child body nodes attached to the joint
   \end{itemize}
\end{enumerate} 

\subsubsection{Haptic interface}

The haptic interface class is a wrapper
that unifies the haptic interface 
creation and attachment processes in SOFA.
The haptic interface builder
contains the following parameters:

\begin{itemize}
  \item \textit{Device}: haptic interface device model (e.g. Omega or Geomagic).
  \item \textit{Scale}: the scale of the haptic interface position data in the scene.
  \item \textit{Origin}: translation and orientation of the haptic interface reference frame.
  \item \textit{Controlled body}: a rigid physical body that will be controlled with this haptic interface. If specified, Filasofia will create a 6-\gls{dof} spring force field connecting the position of the haptic interface and the position of the body.
\end{itemize} 

\subsection{Visual server}

The visual server is responsible for rendering
visual entities on the screen, including
lights, cameras and views, UIs and visual bodies.
The visual server is based on the Filament framework~\cite{filament}.
This is a lightweight \gls{pbr} engine written in C++.
It produces believable graphics without 
a significant overhead in performance costs.
Filament enables the rendering of such \Gls{pbr} phenomena
as ambient occlusion, subsurface scattering,
lens-related and camera-related effects,
and many more. 

Each visual class in the simulation is 
inherited from the \textit{Transform} class,
which contains the translation and orientation
of the object.
Any Transform object can be attached 
to another Transform or mapped to a physical body.
This gives the user a certain level of flexibility 
to build complex structures 
consisting of several entities, 
for instance,
an endoscope, which is a rigid body
with a camera and a light source
attached at the proximal end.

When configuring a visual body,
the user can choose a material 
from the pool of materials already created in the simulation
or create a new one specifically for this body.
In both cases, the user can tune the parameters
of the material (textures, material properties, etc.)
that will be applied exclusively to this body.

\subsubsection{Online subdivision}

\begin{figure}[ht]
\centering
\includegraphics[width=0.45\textwidth]{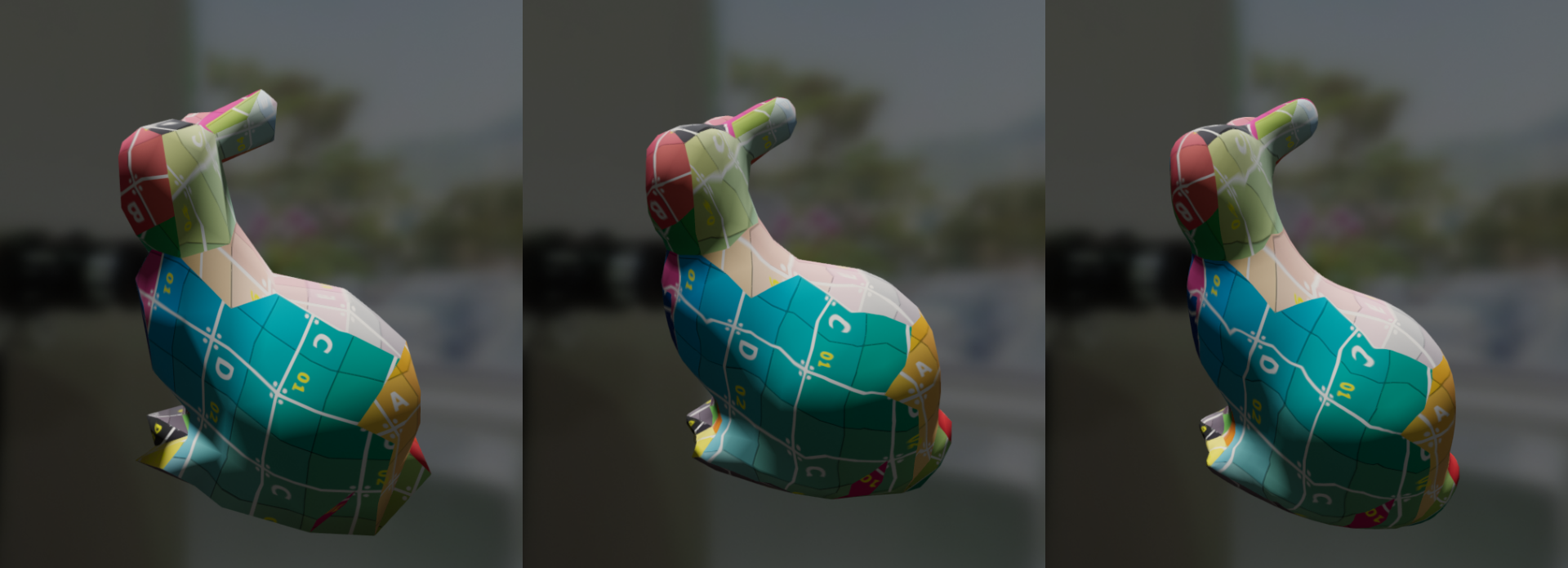}
\caption{Real-time subdivision in Filasofia.
Stanford bunny deformation simulation:
(left) without subdivision,
(center) with level 1 subdivision,
and (right) with level 2 subdivision.}
\label{fig:subdivision}
\end{figure}

If mesh refinement is required, 
a user can activate online subdivision
of the visual mesh (Figure \ref{fig:subdivision}).
In this case, the user should only specify
the required subdivision level 
and the framework will automatically 
introduce new vertices in the topology
to smooth out the surface of the visual body.
This feature is implemented using OpenSubdiv \cite{opensubdiv}.

\subsubsection{User interface}

Filasofia provides a developer 
with a convenient mechanism 
for creating \glspl{gui}.
To facilitate this feature,
the ImGUI library was integrated into the framework \cite{imgui}.
ImGUI is an immediate-mode \gls{gui} library for C++. 
It outputs optimized vertex buffers that can be rendered 
in a 3D-pipeline-enabled application.
Two types of \gls{gui} entities
are available in Filasofia.
First, the \textit{Canvas} class allows
a developer to create a \gls{gui} canvas
and add it to the scene.
A user cannot interact with a canvas
by means of regular input.
However, the canvas can be used
as an indicator or label that can be attached 
to any physical or visual body
like any other visual entity in the scene.
Second, the \textit{Gui} class
is a canvas entity that is attached to a viewport
and overlays on top of the view assigned to this viewport.
Unlike a \textit{Canvas} object, 
a \textit{Gui} accepts user input
and can be used to implement
interaction with the user
by means of buttons, sliders, text fields, 
and other controls.
Both the \textit{Canvas} and \textit{Gui} classes
use the \textit{renderCallback} 
parameter to create \gls{gui} elements.
The render callback can be reassigned in run-time
to enable dynamic \glspl{gui}.

\subsection{Prefabricated entities}

\begin{figure}[ht]
\centering
\includegraphics[width=0.4\textwidth]{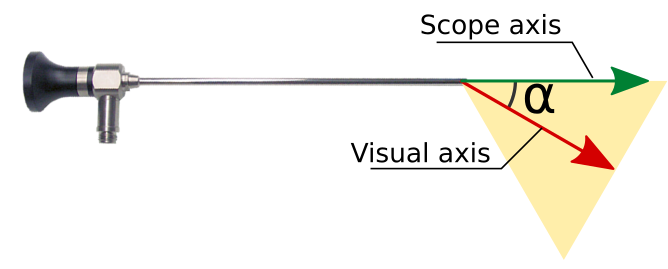}
\caption{The inclination of the visual axis of an endoscope.
The green arrow defines the axis of the endoscope;
the red arrow defines the visual axis.}
\label{fig:endoscope}
\end{figure}

To simplify the process of the creation of 
modalities with common functionality,
Filasofia provides a group of 
prefabricated items
to accelerate the development process.
At the moment, three modalities are 
available in Filasofia.
The first prefab is \textit{ArticulatedInstrument},
which represents a rigid endoscopic instrument
with a 1-\gls{dof} articulated tip
(with one or two moving jaws).
This class can be used to simulate,
for instance, endoscopic forceps or scissors.
The second prefab is \textit{Endoscope},
which is a rigid body with a light source
and a camera attached at the distal end.
The endoscope has two parameters:
the visual axis inclination and the scope roll (Figure~\ref{fig:endoscope}).
The visual axis inclination defines the angle
between the field of view axis and the axis of the endoscope.
The scope roll defines the angle of rotation
of the visual axis around the scope axis.
The third prefab that is currently under development is \textit{MetricTracker}.
The \textit{MetricTracker} class allows the user
to observe the state of a body it is attached to 
and store the records in a dedicated file or container.
This, for instance, can be used to 
track the trajectory of an instrument or
assess the level of force applied to a body.

\section{Performance evaluation}

\begin{figure}[t]
\centering
\includegraphics[width=0.5\textwidth]{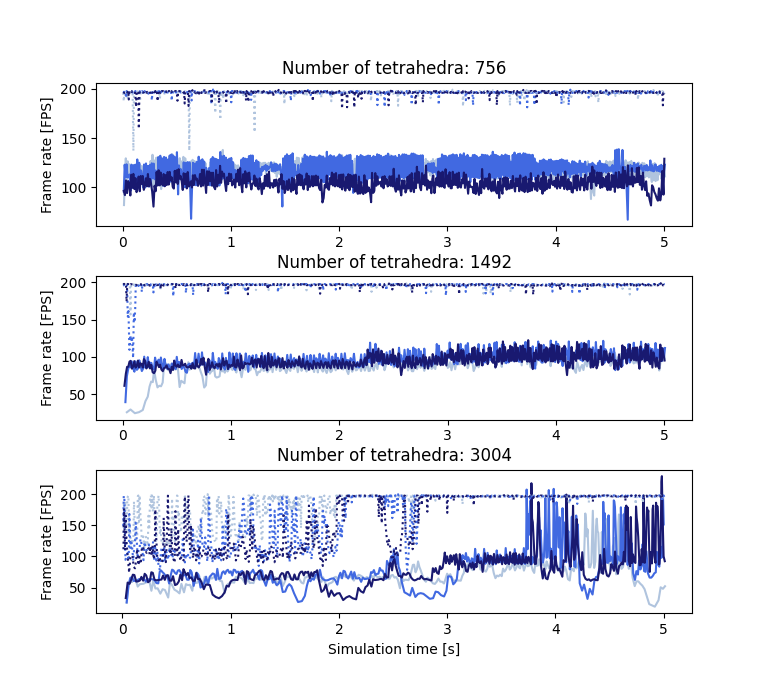}
\caption{Physics update rate comparison 
between SOFA and Filasofia.
The sample scene simulates 
the deformation of the Stanford bunny
using tetrahedron \gls{fe} modelling
with 756 (top), 1492 (middle), 
and 3004 (bottom) tetrahedra.
SOFA simulation framerate
is plotted with solid lines;
Filasofia simulation framerate
is plotted with dotted lines.
Grey colour depicts 
a simulation with a visual model without subdivision;
light blue colour depicts 
a simulation with a visual model with level 1 subdivision;
dark blue colour depicts 
a simulation with a visual model with level 2 subdivision
(using real-time subdivision in Filasofia 
and barycentric mapping in SOFA).
}
\label{fig:benchmarking}
\end{figure}

To evaluate the performance of the framework,
we compared its efficiency against
the standard SOFA implementation
in a simple task of deformation simulation.
An \gls{fe} model of a Stanford bunny is pinned
in space in the head area; 
the rest of the body is free-falling under own weight
(Figure~\ref{fig:subdivision}).
We recorded the physics update rate of 
the same simulation both in SOFA and Filasofia
with two variable parameters:
the \gls{fe} model resolution
and the visual model resolution.
Since SOFA does not feature online subdivision,
a prior subdivision in Blender was used 
for the visual model in the SOFA simulation;
For the Filasofia simulation,
the online subdivision was used instead.

Figure \ref{fig:benchmarking} depicts
the comparison between the same simulation scene
implemented in SOFA and Filasofia.
All tests were executed on a PC
with Intel CORE i5 processor (8th generation)
and NVIDIA Quadro P1000 graphics card.

For the small resolution of the \gls{fe} model (756 elements),
the median frame rate in Filasofia was
$196\pm12$ \gls{fps}, $196\pm3$ \gls{fps}, and $196\pm5$ \gls{fps}
for the visual models with no subdivision, level 1 subdivision, 
and level 2 subdivision, respectively.
For the same \gls{fe} model,
the median frame rate in SOFA was
$117\pm59$ \gls{fps},
$117\pm121$ \gls{fps},
$103\pm57$ \gls{fps}, respectively.

For the medium resolution of the \gls{fe} model (1492 elements),
the median frame rate in Filasofia was
$196\pm3$ \gls{fps}, $196\pm43$ \gls{fps}, and $196\pm2$ \gls{fps}
for the visual models with no subdivision, level 1 subdivision, 
and level 2 subdivision, respectively.
For the same \gls{fe} model,
the median frame rate in SOFA was
$90\pm136$ \gls{fps},
$96\pm95$ \gls{fps},
$93\pm89$ \gls{fps}, respectively.

For the high resolution of the \gls{fe} model (3004 elements),
the median frame rate in Filasofia was
$196\pm431$ \gls{fps}, $196\pm1297$ \gls{fps}, and $196\pm1533$ \gls{fps}
for the visual models with no subdivision, level 1 subdivision, 
and level 2 subdivision, respectively.
For the same \gls{fe} model,
the median frame rate in SOFA was
$67\pm571$ \gls{fps},
$77\pm1256$ \gls{fps},
$76\pm1130$ \gls{fps}, respectively.

It can be seen both from the figure 
and the numerical results that
Filasofia demonstrates a higher framerate
in different resolutions of the mechanical model
even with real-time subdivision enabled.
A higher level of variance in the cases
of high-resolution models can still be mitigated  
with the slack parameter of the physics server.
\section{Sample project}

Using Filasofia, our lab designed
a training platform for in-office hysteroscopy
(Figure \ref{fig:filasofia}).
The simulation platform is designed
for training in spatial navigation 
and instrument manipulation
in basic hysteroscopic tasks:
the passage of the cervical canal 
and the removal of polyps.
The hysteroscope is 
a slender telescope inserted in the uterine cavity
to diagnose and treat intrauterine pathologies.
The hysteroscope is equipped with 
an optical tube that visualizes 
the scene at the distal tip with a 30-degree inclination angle,
a light source,
and a working channel, through which
various hysteroscopic tools, 
e.g. forceps or scissors,
can be deployed.
In the simulation, the task for a trainee
is to enter the uterine cavity through the cervical canal
and remove the polyp 
in the minimum timeframe 
and without applying excessive force.

From the standpoint of implementation,
this simulator showcases most of the previously 
mentioned features.
Both the hysteroscope and 
the hysteroscopic forceps are implemented
using prefabricated items available in Filasofia.
The uterus is a non-dynamic soft \gls{fem} body
with 1702 tetrahedra; 
the visual model of the uterus is configured 
to have the level 1 subdivision for a higher-quality mesh.
The polyp is a dynamic soft \gls{fem} body
that is barycentrically attached to the uterus.
Barycentric attachment allows us
to randomly generate the location of the polyp 
at each restart of the simulation. 
Although cutting tools are not yet available in the simulator,
the polyp can still be removed by tearing it off the uterine wall.
For the basic cases of polypectomy,
it should be sufficient to teach the trainee to perform 
this manoeuvre without risk of perforation. 
\section{Conclusions}

In this work, we presented a framework
for surgical simulation
combining both real-time \gls{fe} modelling 
and \gls{pbr} visuals.
The design approach taken intends to 
simplify the development routine
while serving as a flexible solution
when fine-tuning is required.
Future work will focus on further simplification 
of the development process for the user.
This includes improvements of the build system
to envelop all dependencies in the project,
while still being able to link external versions
of each third-party library;
additional prefabricated content
facilitating the creation of common abstractions,
such as task descriptors and analysers,
guidance tools, and more instrument modalities,
including instrument with additional \glspl{dof} 
and soft instruments.

\bibliographystyle{IEEEtran}
\bibliography{IEEEabrv,bibliography.bib}

\end{document}